\documentclass[journal]{IEEEtran}
\usepackage{graphicx,amssymb,amsmath}
\usepackage[noadjust]{cite}
\usepackage{setspace}               
\usepackage{stfloats}
\usepackage{bbding}
\usepackage{amsthm}
\usepackage{amsmath}
\usepackage{cases,subeqnarray}
\usepackage{bm,multirow,bigstrut}
\usepackage{epstopdf}
\usepackage{textcomp}
\usepackage{latexsym,bm}
\usepackage{booktabs,changebar}
\usepackage{xcolor}
\usepackage{mathtools}
\usepackage{dsfont}
\usepackage{extarrows}
\usepackage{mathrsfs}
\usepackage{subfigure}

\usepackage{cite}
\usepackage{bm}
\usepackage{cleveref}
\usepackage{multicol}       
\usepackage{multirow}       
\usepackage{array}          
\usepackage{colortbl}
\usepackage{makecell}
\definecolor{crimson}{RGB}{192,0,0}         
\definecolor{navy}{RGB}{47,85,151}         

\makeatletter                               
\newif\if@restonecol
\makeatother

\makeatletter
\newif\if@restonecol
\makeatother

\usepackage[linesnumbered,ruled,vlined]{algorithm2e}
\usepackage{algpseudocode}
\usepackage{amsmath}
\theoremstyle{plain}

\theoremstyle{plain}

\IEEEoverridecommandlockouts

\begin{document}
\title{Joint Power Control and Precoding for Cell-Free Massive MIMO Systems With Sparse Multi-Dimensional Graph Neural Networks}
\author{{Yukun~Ma, Jiayi~Zhang,~\IEEEmembership{Senior Member,~IEEE}, Ziheng~Liu, Guowei Shi, and Bo~Ai,~\IEEEmembership{Fellow,~IEEE}}
\thanks{Y. Ma, J. Zhang, Z. Liu, and B. Ai are with the School of Electronic and Information Engineering and also with the Frontiers Science Center for Smart High-Speed Railway System, Beijing Jiaotong University, Beijing 100044, China (e-mail: \{21721018, jiayizhang, zihengliu, boai\}@bjtu.edu.cn). Guowei Shi is with China Academy of Information and Communications Technology, Beijing, China, 100083. (email: shigw@sohu.com)}
}
\maketitle
\begin{abstract}
Cell-free massive multiple-input multiple-output (CF mMIMO) has emerged as a prominent candidate for future networks due to its ability to significantly enhance spectral efficiency by eliminating inter-cell interference. However, its practical deployment faces considerable challenges, such as high computational complexity and the optimization of its complex processing. To address these challenges, this correspondence proposes a framework based on a sparse multi-dimensional graph neural network (SP-MDGNN), which sparsifies the connections between access points (APs) and user equipments (UEs) to significantly reduce computational complexity while maintaining high performance. In addition, the weighted minimum mean square error (WMMSE) algorithm is introduced as a comparative method to further analyze the trade-off between performance and complexity. Simulation results demonstrate that the sparse method achieves an optimal balance between performance and complexity, significantly reducing the computational complexity of the original MDGNN method while incurring only a slight performance degradation, providing insights for the practical deployment of CF mMIMO systems in large-scale network.
\end{abstract}


\section{Introduction}
Wireless communications technology has consistently evolved to address the increasing demands for higher data rates, enhanced reliability, and efficient resource utilization \cite{he2021cell}. In this context, cell-free massive multiple-input multiple-output (CF mMIMO) has been recognized as a disruptive approach to meet the performance requirements of beyond fifth-generation (B5G) and sixth-generation (6G) networks \cite{chen2020structured}. By adopting a distributed architecture with numerous low-cost and spatially dispersed access points (APs), CF mMIMO facilitates joint servicing of user equipments (UEs) over shared frequency-time resources, effectively mitigating inter-cell interference and redefining the concept of network boundaries.

The advancement of B5G wireless networks demands higher spectral efficiency (SE) and scalability to accommodate increasing densities of user devices and diverse applications. While CF mMIMO systems present a compelling solution by utilizing distributed APs to jointly serve UEs without inter-cell interference, their deployment faces significant challenges, particularly in the areas of resource management in wireless communications \cite{zhang2025multi}. Managing power allocation, precoding, and beamforming in such networks is particularly complex and resource-intensive. Traditional optimization methods, such as the weighted minimum mean square error (WMMSE) algorithm, are effective but computationally expensive \cite{fu2024wmmse}. As the number of antennas in CF mMIMO systems grows, these approaches struggle to maintain practicality, highlighting the need for scalable and efficient alternatives \cite{chen2025channel,zhang2025enhancing,zhang2024enhancing}. For instance, recent works utilize Rate-Splitting Multiple Access (RSMA) alongside optimization methods to enhance uplink performance with low-resolution ADCs \cite{zhang2025enhancing} or to improve downlink secrecy under hardware impairments \cite{zhang2024enhancing}, reflecting ongoing advancements in resource allocation.

For these challenges, data-driven approaches based on deep learning have gained significant attention in recent years. Among them, graph neural networks (GNNs) have emerged as a standout solution due to their ability to model the complex relationships between APs, UEs, and channel signal information (CSI) in wireless networks as graph structures \cite{shen2022graph,mishra2024graph,yang2020graph}. In \cite{mishra2024graph}, permutation equivariance and attention mechanisms are leveraged to achieve efficient power control across diverse network scales. Additionally, \cite{yang2020graph} describes channel spatial correlations using graph structures and captures temporal correlations by combining time-adjacent graphs, achieving superior performance. By leveraging the inherent graphical properties of wireless networks, GNNs provide a scalable and efficient framework for tackling these problems. Nevertheless, the practical application of GNNs in CF mMIMO systems is not without its limitations \cite{liu2024graph}. The performance gap between GNN-based solutions and traditional optimization methods remains an open research question, especially in achieving near-optimal SE. 

Traditional GNNs often struggle with information loss during message passing due to dimensionality compression, limiting their ability to capture the multi-dimensional nature of problems like CF mMIMO. Multi-dimensional graph neural networks (MDGNNs) \cite{liu2023multidimensional} address this by updating the hidden representations of hyper-edges and leverages permutation priors to enhance learning efficiency and preserving the full dimensionality of input and output tensors, which avoids dimension compression. This allows to better exploit permutation invariance and multi-dimensional relationships, making them ideal for complex wireless communication scenarios.

\begin{figure}[t]
    \centering
    \includegraphics[width=1\linewidth]{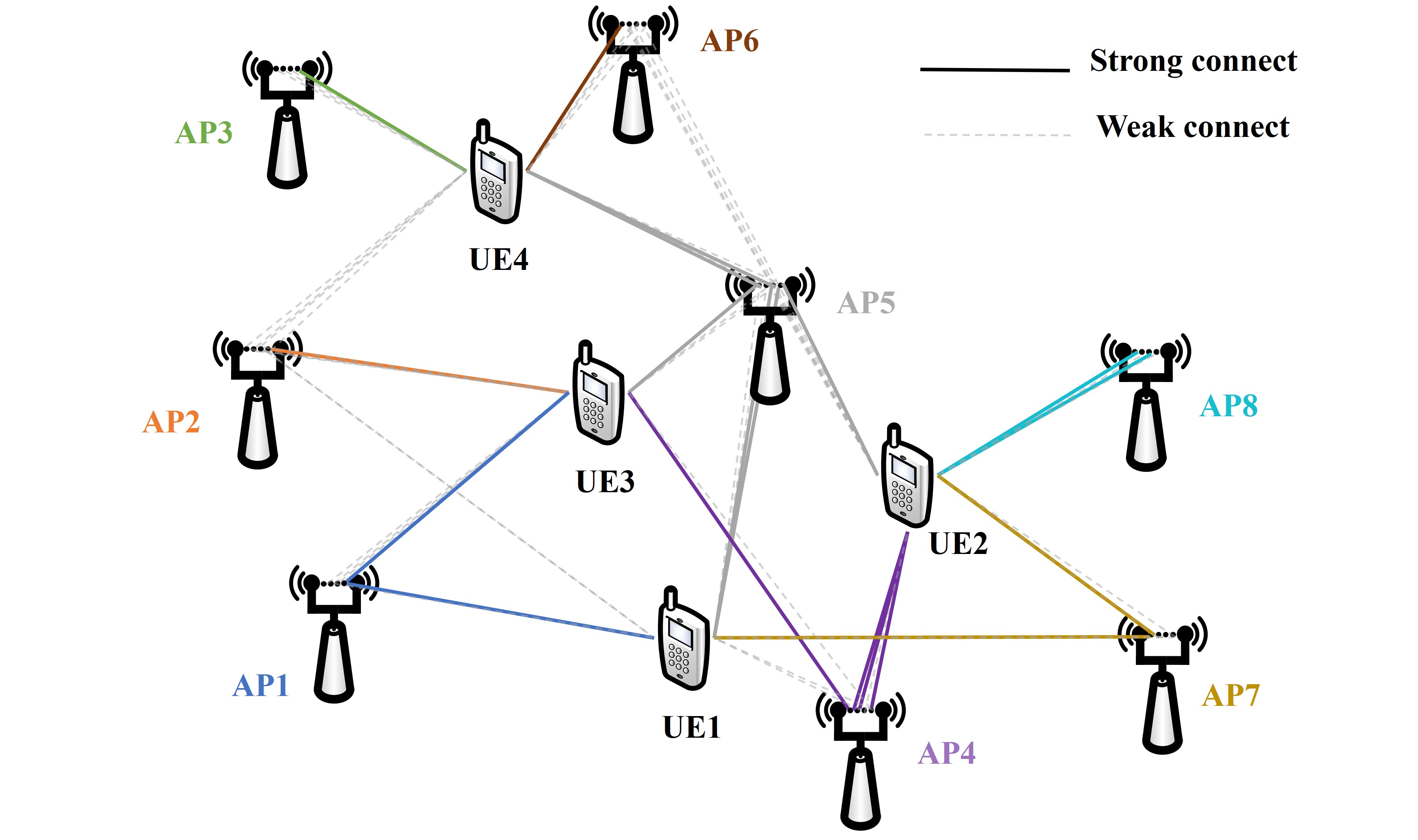}
    \caption{AP-UE connections in SP-MDGNN. Illustrating APs with multiple antennas efficiently serving UEs over a range, where links are labeled ``strong" (solid lines) or ``weak" (dashed lines) based on channel strength, with weaker connections strategically trimmed adopting the sparse method to significantly lower complexity in large-scale CF mMIMO networks.}
    \label{fig:AP-UE connections}
\end{figure}

Although MDGNNs method has lower complexity, potential computational bottlenecks may still arise in future CF mMIMO systems with huge antennas \cite{liu2023double}. their computational complexity is still high. To tackle these problems, especially for power control and precoding, we propose SP-MDGNN that employs a sparse frame to MDGNN significantly reduce complexity while maintaining performance with only a marginal trade-off. To illustrate the sparse connectivity structure in power control, as shown in Fig. \ref{fig:AP-UE connections}, each AP equipped with multiple antennas, serves UEs within a distance threshold. Connections are categorized as ``strong" or ``weak", with weak connections pruned based on the sparse adjacency matrix. The main contributions are given as follows: 

\begin{enumerate}
    \item We address the challenge of high computational complexity in large-scale CF mMIMO systems by proposing the SP-MDGNN framework, which hugely reduces the complexity through pruning low-impact connections while essentially maintaining SE, making it promising for resource management in wireless communications.
    \item We tackle the issue of optimizing AP-UE connections in MDGNNs by implementing attention mechanisms that assess the importance of these connections, thereby improving SE while ensuring scalability.
    \item We investigate the performance-complexity trade-off in CF mMIMO systems through simulations conducted under varying sparse thresholds, thereby identifying an optimal balance that enhances the efficiency of next-generation wireless networks.
\end{enumerate}

\section{System Model}
In this correspondence, we investigate a CF mMIMO system consisting of $L$ APs, each equipped with $N$ antennas, and $K$ single-antenna UEs. These entities are randomly distributed within a two-dimensional square area of side length $D$ meters. The channel characteristics between APs and UEs are captured by the channel matrix $\mathbf{H} \in \mathbb{C}^{L \times K \times N}$, where the element $\mathbf{H}_{j,k,n}$ represents the complex channel gain from the $n$-th antenna of the $j$-th AP to the $k$-th UE. The channel gain is modeled as:

\begin{equation}
    \mathbf{H}_{j,k,n}=\sqrt{\beta_{j,k}}\mathbf{R}_{j,k}^{1/2}\mathbf{h}_{j,k,n},
\end{equation}
where $\beta_{j,k}$ denotes the large-scale fading coefficient, accounting for path loss and shadowing effects between the $j$-th AP and the $k$-th UE. The term $\mathbf{H}_{j,k,n}$ represents the small-scale fading coefficient, modeled as an independent and identically distributed (i.i.d.) complex Gaussian random variable, i.e., $\mathbf{H}_{j,k,n} \sim \mathcal{CN}(0,1)$. The spatial correlation matrix $\mathbf{R}_{j,k} \in \mathbb{C}^{N \times N}$ captures the antenna array’s spatial correlation at the $j$-th AP with respect to the $k$-th UE.

Specifically, each AP transmits signals via the transmission matrix $\mathbf{F} \in \mathbb{C}^{L \times K \times N}$, where the vector $\mathbf{F}_{j,k} \in \mathbb{C}^{N}$ represents the transmission from AP $j$ to UE $k$. The matrix $\mathbf{F}$ jointly determines power control and precoding: the norm $\|\mathbf{F}_{j,k}\|^2$ dictates the allocated power, while the complex vector $\mathbf{F}_{j,k}$ itself defines the beamforming direction. Optimizing $\mathbf{F}$ aims to maximize SE by jointly finding the optimal power levels and beam directions, subject to the per-AP power constraint:

\begin{equation}
    \sum_{k=1}^K \|\mathbf{F}_{j,k}\|^2 \leq P_{\text{max}}, \quad \forall j \in \{1, 2, \dots, L\},
\end{equation}
where $\mathbf{F}_{j,k} \in \mathbb{C}^{N}$ is the transmission vector from the $j$-th AP to the $k$-th UE, and $\|\cdot\|^2$ denotes the squared Euclidean norm. This constraint ensures that the power allocation remains feasible within the system’s capacity.

The received signal $y_k$ at the $k$-th UE, is expressed as:

\begin{equation}
    y_k = \sum_{j=1}^{L} \mathbf{h}_{j,k}^H \mathbf{F}_{j,k} s_k + \sum_{j=1}^{L} \sum_{i \neq k}^{K} \mathbf{h}_{j,k}^H \mathbf{F}_{j,i} s_i + n_k,
    \label{equ1}
\end{equation}
where $s_k$ is the transmitted symbol for the $k$-th UE, with $\mathbb{E}[|s_k|^2] = 1$, and $n_k \sim \mathcal{CN}(0, \sigma^2)$ represents additive white Gaussian noise (AWGN) with variance $\sigma^2$. Furthermore, the first term in (3) corresponds to the desired signal, while the second term captures inter-user interference from other UEs.

The signal-to-interference-plus-noise ratio (SINR) for the $k$-th UE is defined as:

\begin{equation}
    \text{SINR}_k = \frac{\left| \sum_{j=1}^{L} \mathbf{h}_{j,k}^H \mathbf{F}_{j,k} \right|^2}{\sum_{i \neq k}^{K} \left| \sum_{j=1}^{L} \mathbf{h}_{j,k}^H \mathbf{F}_{j,i} \right|^2 + \sigma^2},
\end{equation}
where $\mathbf{h}_{j,k} \in \mathbb{C}^{N}$ is the channel vector from the $j$-th AP to the $k$-th UE, and the superscript $H$ denotes the Hermitian transpose. This formulation quantifies the quality of the received signal by balancing the desired signal power against interference and noise.

The SE for the $k$-th UE is then given by:
\begin{equation}
    \text{SE}_k = \log_2 (1 + \text{SINR}_k).
\end{equation}

Consequently, the total SE of the system is the sum of individual SEs across all UEs, i.e., $\sum_{k=1}^{K} \text{SE}_k$. Moreover, the primary objective of this work is to optimize the transmission matrix $\mathbf{F}$ to maximize the system’s total SE. Specifically, the goal is formulated as follows:

\begin{equation}
    \begin{aligned}
        \underset{\{\mathbf{F}_{j,k}\}}{\text{maximize}} \quad 
         \sum_{k=1}^K & \log_2(1 + \text{SINR}_k), \\
        \text{subject to}  \quad 
         \sum_{k=1}^K \|\mathbf{F}_{j,k}\|^2  \leq & P_{\text{max}},  \quad \forall j \in \{1, 2, \dots, L\}.
    \end{aligned}
    \label{eq:nonconvex}
\end{equation}

This optimization problem is non-convex and NP-hard due to the coupling of variables in the SINR expression. Conventional methods often transform it into a convex form and solve it iteratively, but their computational complexity grows significantly with network size. To address this, we propose SP-MDGNN to achieve low-complexity solutions while maintaining near-optimal SE, as detailed in subsequent sections.

\section{Proposed SP-MDGNN Method}
We model the CF mMIMO system as a graph, enabling MDGNNs to capture and optimize the intricate channel relationships between APs and UEs. In this graph, nodes represent APs and UEs, and edges signify channel connections weighted by CSI. The MDGNN processes this information via message passing. Our SP-MDGNN extends this by providing a unified framework for the coupled tasks of power control and precoding. While distinct network pathways may primarily address power allocation scaling and precoding direction, the entire SP-MDGNN is trained end-to-end with a singular objective: maximizing overall system SE. This joint training implicitly learns the optimal interplay in coupled strategies.

To enhance computational efficiency, we introduce sparse technology to dynamically prune low-impact connections. Recent works, such as \cite{wang2023sparse}, have introduced a sparse graph isomorphism network method using $l_0$-norm regularization to prune interference links, reducing CSI overhead while maintaining performance. While \cite{wang2023sparse} targets two-timescale resource allocation in interference channels, our SP-MDGNN focuses on CF mMIMO, optimizing power control and precoding with a different sparsification approach.

\subsection{Sparse Adjacency Matrix Design}
Given the channel matrix $\mathbf{H}$, we begin by defining a learnable adjacency tensor $\mathbf{A}$, which is updated via a trainable parameter $\mathbf{W}$ through a sigmoid activation function:
\begin{equation}
   A_{j,k,n} = \sigma(W_{j,k,n}), 
\end{equation}
where $\sigma(x) = \frac{1}{1 + e^{-x}}$ is the sigmoid function, and $\mathbf{W}$ is the trainable weight tensor.

To enforce sparsity, a threshold $\tau$ is applied to the adjacency tensor to generate a sparse mask $\mathbf{M}$:
\begin{equation}
    M_{j,k,n} =
\begin{cases}
1, & \text{if } A_{j,k,n} > \tau, \\
0, & \text{otherwise}.
\end{cases}
\end{equation}

Nonetheless, the sparse channel matrix $\mathbf{H}_\text{sparse}$ is then computed as the element-wise product of the original channel matrix $\mathbf{H}$ and the sparse mask $\mathbf{M}$:

\begin{equation}
   \mathbf{H}_\text{sparse} = \mathbf{H} \odot \mathbf{M}, 
\end{equation}
where $\odot$ denotes the Hadamard product, pruning low-impact connections while retaining critical connections.

The sparsification mechanism, using a learnable adjacency tensor, is universal in SP-MDGNN, pruning low-impact connections to reduce GNN complexity. Although consistent across tasks, the input feature dimensions and optimization objectives differ. In power control, sparsification targets AP-UE link features to optimize power allocation among users, whereas in precoding, it focuses on user-antenna link features to enhance beamforming and interference management. 

\begin{figure}[t]
    \centering
    \includegraphics[width=1\linewidth]{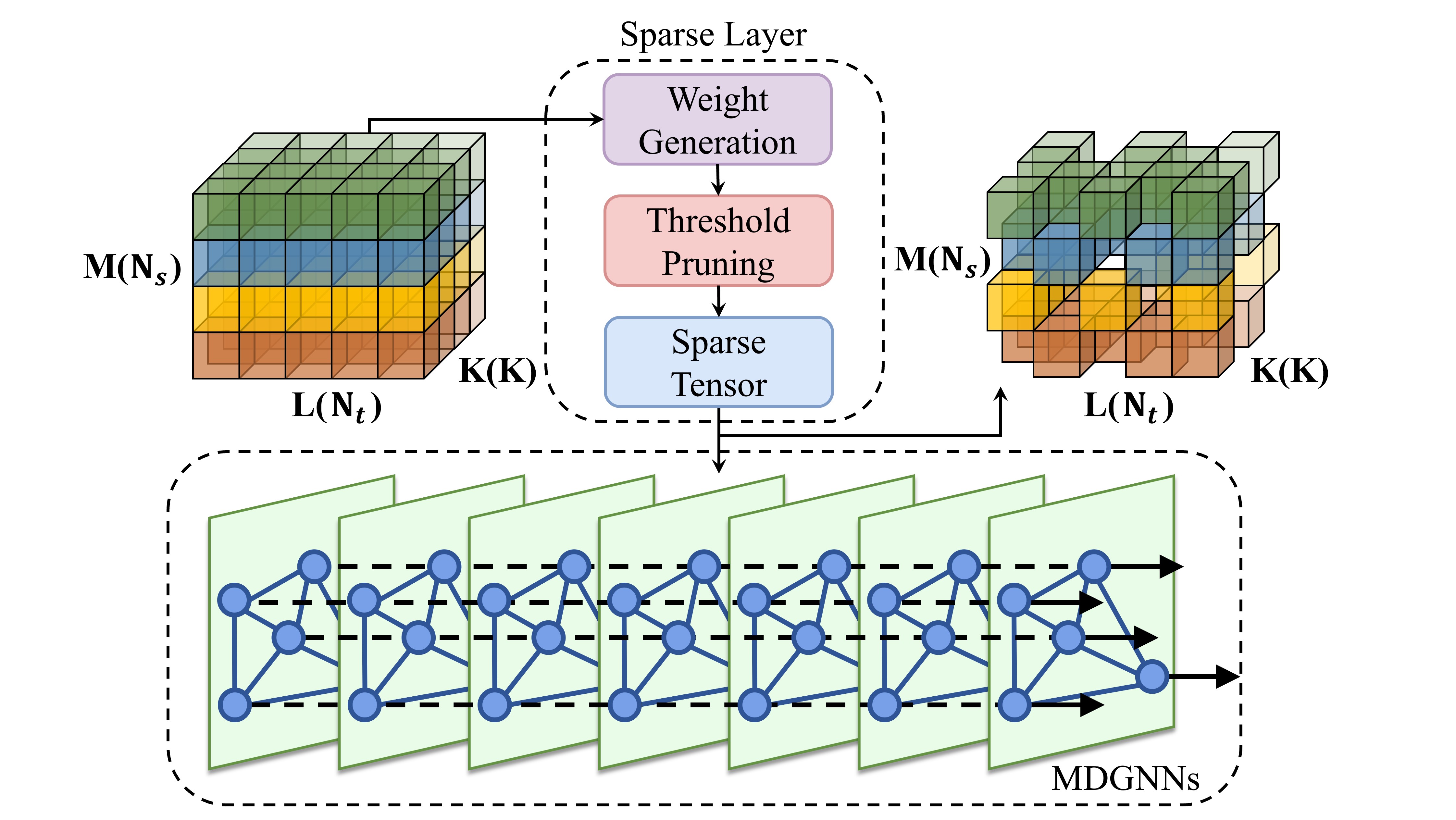}
    \caption{Framework of the proposed SP-MDGNN. The dimensions shown illustrate the  adaptability to different tasks: In power control, dimensions relate to L (APs), K (UEs), M (antennas), and sparsity focuses on AP-UE links. In precoding, dimensions relate to K (UEs), N$_t$ (antennas), N$_s$ (RF chains), and sparsity focuses on user-antenna links.}
    \label{Sparse MDGNNs}
\end{figure}

\subsection{GNN Layer Propagation}
The sparse channel matrix $\mathbf{H}_\text{sparse}$ serves as input to the GNN with multiple layers. For the $l$-th layer, the output is:
\begin{equation}
\mathbf{X}^{(l)} = \phi \left( \sum_{i=1}^{4} \mathbf{P}_i^{(l)} \cdot \left( \mathbf{A} \odot \mathbf{X}^{(l-1)} \right) \right),
\label{eq:gnn_layer}
\end{equation}
where $\mathbf{X}^{(l-1)}$ is the input feature tensor, $\mathbf{P}_i^{(l)}$ is the learnable parameter matrix for the $i$-th aggregation, $\mathbf{A}$ is the sparse adjacency matrix that restricts aggregation to significant links only, and $\phi(\cdot)$ is the activation function. The Hadamard product $\odot$ ensures computations focus on unpruned connections, minimizing complexity in large-scale CF mMIMO networks.

The SP-MDGNN framework minimizes computations by pruning low-impact connections in large-scale CF mMIMO networks, enhancing efficiency while maintaining high performance, as demonstrated in the following experimental results.

\subsection{Output Layer Design}
The output of the SP-MDGNN is used to compute the transmission matrix $\mathbf{F}$. For power control, the network output is normalized to satisfy the per-AP power constraint:
\begin{equation}
    \mathbf{F}_{j,k} = \frac{\mathbf{X}_{j,k}^{(L)}}{\|\mathbf{X}_{j,k}^{(L)}\|} \sqrt{\frac{P_\text{max}}{\sum_{k=1}^K \|\mathbf{X}_{j,k}^{(L)}\|^2}},
\end{equation}
where $\mathbf{X}_{j,k}^{(L)}$ is the output from the $j$-th AP to the $k$-th UE, determining both power allocation magnitude and beamforming direction. For precoding tasks, the output $\mathbf{X}^{(L)}$ directly forms the precoding vectors, leveraging the learned sparse channel features to maximize SINR through optimal beam design.

\subsection{Joint Power Control and Precoding Optimization}
The SP-MDGNN employs a unified multi-task framework to jointly optimize power control and precoding. The network consists of shared sparse feature extraction layers followed by task-specific output heads. Both tasks utilize the same sparse channel representation $\mathbf{H}_{\text{sparse}}$, enabling to capture the inherent coupling between power allocation and beamforming.

The joint optimization is achieved through an end-to-end training process with a combined loss function:
\begin{equation}
    \mathcal{L}_{\text{joint}} = \alpha \mathcal{L}_{\text{power}} + (1-\alpha) \mathcal{L}_{\text{prec}}.
\end{equation}
where $\mathcal{L}_{\text{power}}$ and $\mathcal{L}_{\text{prec}}$ are the individual task losses, and $\alpha \in [0,1]$ balances their contributions. Both losses aim to maximize SE but from different perspectives: power control optimizes resource allocation while precoding optimizes spatial processing. The shared sparse features and joint training enable the network to implicitly learn the optimal trade-offs between these coupled strategies, resulting in superior overall system performance compared to separate optimization.

Moreover, as shown in Fig. ~\ref{Sparse MDGNNs}, the SP-MDGNN framework utilizes weight generation, threshold pruning, and sparse adjacency to reduce redundant connections, lowering complexity while ensuring accuracy and scalability in CF mMIMO networks both for power control and precoding.

\section{Simulation Results and Analysis}
\subsection{Simulation Environment}
We evaluate the SP-MDGNN framework for power control and precoding in a CF mMIMO network, assuming perfect instantaneous CSI to isolate channel estimation errors and focus on the algorithm’s effectiveness. For power control, $L=9$ APs ($M=8$ antennas) and $K=8$ UEs are randomly distributed in a $1000 \times 1000 \text{m}^2$ area, with channels modeled by Rayleigh fading and large-scale fading $\beta_{j,k} = -30.5 - 36.7 \log_{10}(d_{j,k})$, where $d_{j,k} = \sqrt{10^2 + |\text{AP location} - \text{UE location}|^2}$. APs transmit at $1000 \text{mW}$ with $5 \text{W}$ static power. For precoding, $K=4$ UEs and $N_t=16$ antennas use Rayleigh fading channels, modeled by the SV model with $N_{\text{cl}}=4$ clusters, $N_{\text{ray}}=5$ rays, and a $10^\circ$ angular spread for AoDs. Datasets include 10,000 training and 2,000 testing samples. SP-MDGNN, with 1 sparse layer and 5 GNN layers (256 hidden units), prunes weak links via threshold $\tau$, and is benchmarked against WMMSE and baseline MDGNN for SE and complexity.

\subsection{Trade-off for Separate Precoding and Power Control}
\begin{figure}[t]
    \centering
    \includegraphics[width=1\linewidth]{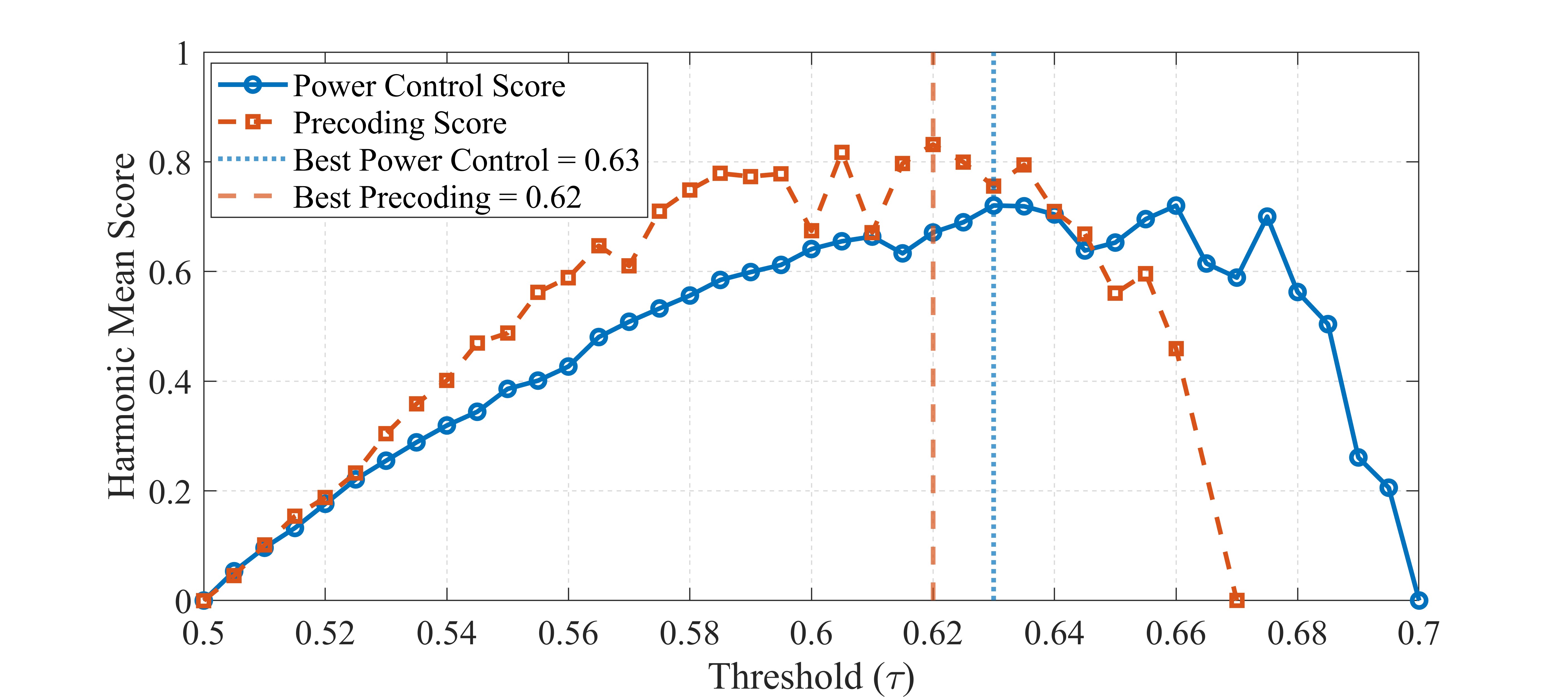}
    \caption{Harmonic mean score selection under different thresholds ($\tau$).}
    \label{Threshold and performance metrics}
\end{figure}

To determine suitable sparsity thresholds ($\tau$) for SP-MDGNN and understand the inherent performance-complexity trade-off, we first conducted evaluations on the separate power control and precoding tasks, varying $\tau$ from 0.5 to 0.7. Our analysis revealed that as $\tau$ increases, the \textit{Sparsity} level, which reflects complexity reduction, generally rises, while \textit{Performance Retention}, which relatives to the non-sparse baseline, tends to decrease, highlighting the trade-off. To quantitatively identify the optimal balance, we defined normalized metrics for \textit{Sparsity} (S) and \textit{Performance Retention} (P), ensuring higher values are better for both. We then calculated their harmonic mean using the formula $H_s = \frac{2 \cdot S \cdot P}{S + P}$ as a unified score. As shown in Fig. \ref{Threshold and performance metrics}, evaluating this $H_s$ score across the tested $\tau$ range for power control and precoding each task revealed distinct peaks at the optimal thresholds providing the best compromise, specifically $\tau_{pc} = 0.63$ for power control and $\tau_{prec} = 0.62$ for precoding. These optimal values, determined from the separate task analyses, were subsequently employed in joint power control and precoding evaluations.

In training, SP-MDGNN converges faster and requires fewer training cycles than baseline MDGNN. By focusing layer updates on the parameters of critical connections, SP-MDGNN uses sparsity as an effective regularization mechanism, suppressing overfitting of minor features such as weak connections, and improving optimization stability. This demonstrates the advantage of sparsity in accelerating training.

To rigorously evaluate the proposed SP-MDGNN framework's efficiency gains, we first compare its performance and computational complexity against the baseline MDGNN when addressing the power control and precoding tasks individually, using the optimal sparsity thresholds determined previously. For power control, SP-MDGNN incurs only a minor 1.3\% decrease in SE while achieving a substantial 55\% reduction in computational complexity. Similarly, for precoding, the performance decrease is merely 1.44\%, accompanied by a 49\% complexity reduction. These results clearly show SP-MDGNN's ability to drastically cut computational costs with minimal impact on performance for the separate tasks.

\subsection{Results Analysis for Joint Precoding and Power Control}
After determining the optimal sparsity thresholds for SP-MDGNN in the separate power control and precoding, we now evaluate its performance and complexity for joint power control and precoding. This evaluation utilizes similar system parameters and consistent channel environment as detailed in the joint setting description of Sec. IV-A. Our implementation employs a multi-task SP-MDGNN framework with dedicated power control and precoding sub-modules sharing the same core architecture and related channel inputs. 
Both sub-modules are trained jointly by minimizing a combined loss function, allowing the model to implicitly capture the interplay between these coupled tasks. We compare it with three methods under identical conditions: the baseline MDGNN, the attention-based MDGNN (A-MDGNN), and the WMMSE algorithm.

\begin{figure}[t]
    \centering
    \includegraphics[width=1\linewidth]{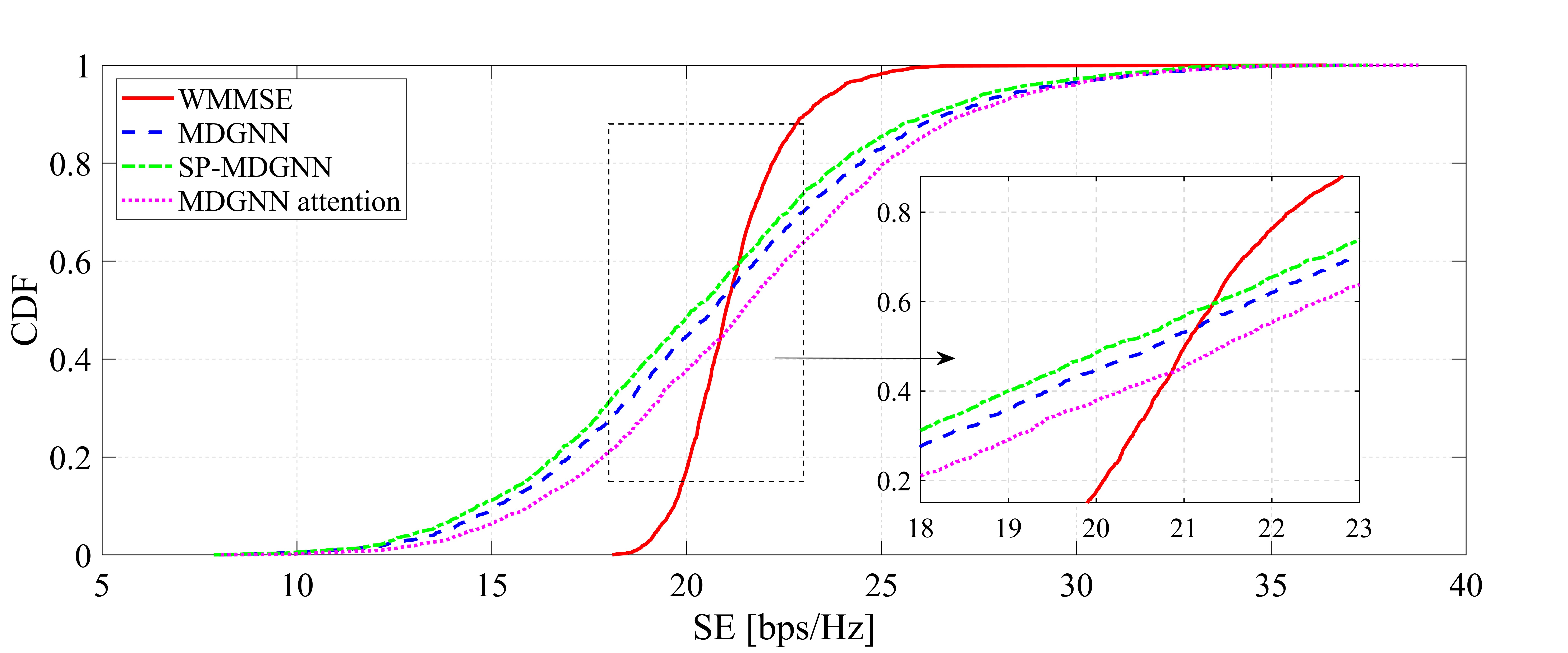}
    \caption{CDF of SE for different joint power control and precoding schemes.}
    \label{CDF of 4 methods comparison}
\end{figure}

The cumulative distribution function (CDF) in Fig. \ref{CDF of 4 methods comparison} compares the SE performance of the four methods for joint power control and precoding. WMMSE achieves the highest mean SE with a concentrated distribution, indicating greater stability. In contrast, MDGNN-based methods show a wider distribution, reflecting higher performance variability. The baseline MDGNN closely matches WMMSE, while SP-MDGNN maintains a competitive SE performance. The attention-based MDGNN performs near WMMSE by attention mechanisms, enhancing SE performance. These results highlight the trade-off among approaches, with the SP-MDGNN method offering flexible solutions across diverse scenarios.

\begin{figure}[t]
    \centering
    \includegraphics[width=1\linewidth]{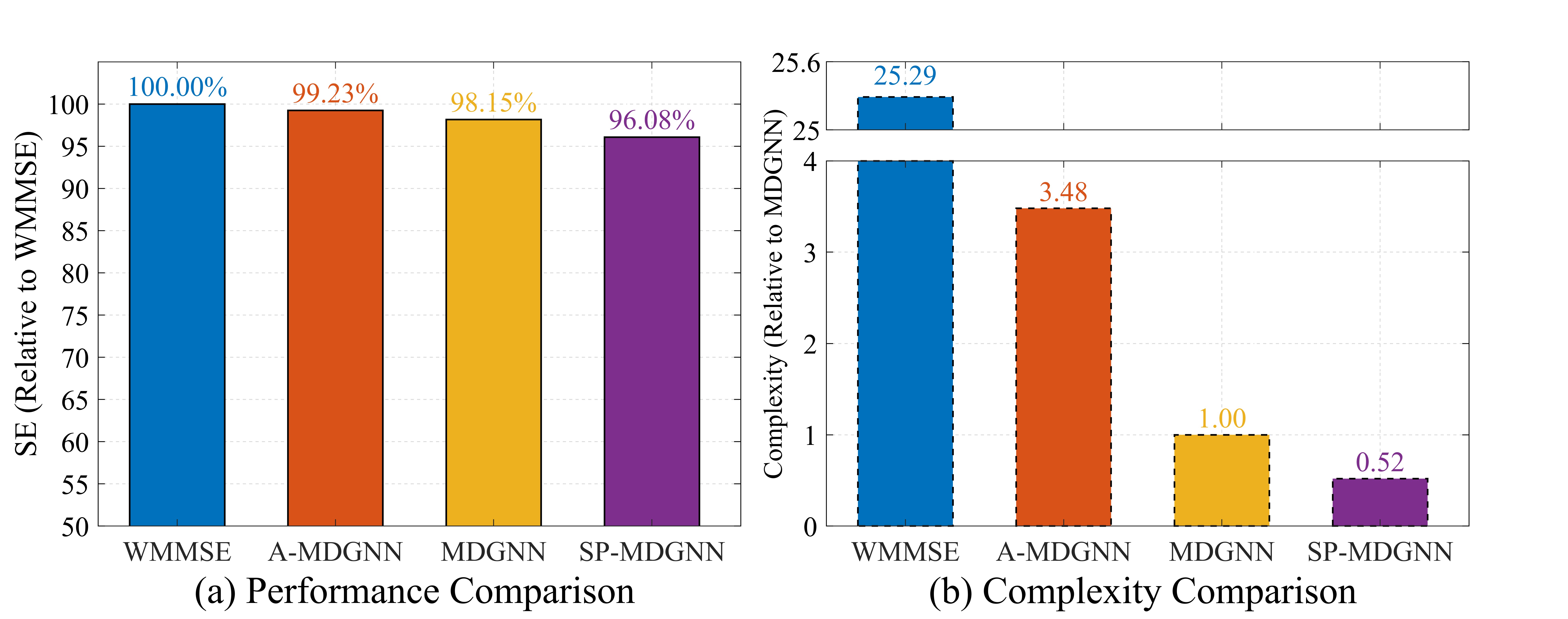}
    \caption{SE Performance and computational complexity comparison for joint power control and precoding in CF mMIMO.}
    \label{Sum rate and complexity of 4 methods}
\end{figure}

The computational complexity comparison presented here is based on measuring the computation time required to process an identical number of samples using the same hardware setup for each method. Fig. \ref{Sum rate and complexity of 4 methods} illustrates the resulting performance-complexity trade-off. While WMMSE offers the highest SE, it incurs substantial computational cost, and the attention-based MDGNN also increases complexity for a slight performance gain over the baseline. The crucial advantage of the proposed SP-MDGNN is highlighted when compared to the original MDGNN: SP-MDGNN experiences only a minor 2.11\% decrease in SE, but achieves a significant 48\% reduction in computational complexity. This demonstrates SP-MDGNN's effectiveness in balancing near-optimal performance with significantly reduced computational demands, making it highly suitable for practical large-scale CF mMIMO deployments.

\section{Conclusion}
This correspondence addresses the challenges of high computational complexity and resource allocation optimization in CF mMIMO by proposing the SP-MDGNN framework. By sparsifying connections, the approach significantly reduces complexity while maintaining near-optimal performance in power control and precoding tasks. The SP-MDGNN achieves substantial complexity reduction with minimal performance loss, making them ideal for deployments of CF mMIMO systems with extremely larger numbers of antennas in the future. Comparative analysis with attention-based MDGNN and WMMSE highlights the trade-off among different methods. Observations during experiments indicate that the proposed method effectively balances SE performance and computational complexity, and notably, exhibits fast convergence characteristics, likely due to the reduced model complexity and focused gradient updates resulting from sparsity. Furthermore, the robustness of the framework is validated across a myriad of network conditions. In future work, we will investigate SP-MDGNN’s dynamic resource allocation in real-time to improve scalability while addressing practical channel estimation challenges under imperfect CSI.
\bibliographystyle{IEEEtran}
\bibliography{IEEEabrv,Ref}
\end{document}